\documentclass[aps,twocolumn,superscriptaddress,nofootinbib]{revtex4-1}

\usepackage[dvips]{graphicx}

\usepackage{amssymb,amsfonts,amsmath}
\usepackage[usenames,dvipsnames]{xcolor}
\usepackage{ulem}  %unter- (\uline{important}) und durchstreichen (\sout{wrong})
\usepackage{booktabs}
\usepackage[english]{babel}

\begin{document}

%% For titles, only capitalize the first letter
%% \title{Almost sharp fronts for the surface quasi-geostrophic equation}

\title{Atomistic study of energy funneling in the light-harvesting complex of green sulfur bacteria}

%% Enter authors via the \author command.
%% Use \affil to define affiliations.
%% (Leave no spaces between author name and \affil command)

%% Note that the \thanks{} command has been disabled in favor of
%% a generic, reserved space for PNAS publication footnotes.

%% \author{<author name>
%% \affil{<number>}{<Institution>}} One number for each institution.
%% The same number should be used for authors that
%% are affiliated with the same institution, after the first time
%% only the number is needed, ie, \affil{number}{text}, \affil{number}{}
%% Then, before last author ...
%% \and
%% \author{<author name>
%% \affil{<number>}{}}

%% For example, assuming Garcia and Sonnery are both affiliated with
%% Universidad de Murcia:
%% \author{Roberta Graff\affil{1}{University of Cambridge, Cambridge,
%% United Kingdom},
%% Javier de Ruiz Garcia\affil{2}{Universidad de Murcia, Bioquimica y Biologia
%% Molecular, Murcia, Spain}, \and Franklin Sonnery\affil{2}{}}

%\author{Joonsuk Huh\affil{1}{Department of Chemistry and Chemical Biology, Harvard University, Cambridge, Massachusetts 02138, United States}, Semion K. Saikin\affil{1}{}, Jennifer C. Brookes\affil{1}{}\affil{2}{Department of Physics and Astronomy, University College London, Gower, London WC1E 6BT}, St\'ephanie Valleau\affil{1}{}, Takatoshi Fujita\affil{1}{} \and Al\'an Aspuru-Guzik\affil{1}{}\thanks{To whom correspondence should be addressed. E-mail: aspuru@chemistry.harvard.edu}
%}
\author{Joonsuk Huh}
\email{Email: huh@fas.harvard.edu}
\affiliation{Department of Chemistry and Chemical Biology, Harvard University, Cambridge, Massachusetts 02138, United States}
\author{Semion K. Saikin}
\affiliation{Department of Chemistry and Chemical Biology, Harvard University, Cambridge, Massachusetts 02138, United States}
\author{Jennifer C. Brookes}
\affiliation{Department of Chemistry and Chemical Biology, Harvard University, Cambridge, Massachusetts 02138, United States}
\affiliation{Department of Physics and Astronomy, University College London, Gower, London WC1E 6BT}
\author{St\'ephanie Valleau}
\affiliation{Department of Chemistry and Chemical Biology, Harvard University, Cambridge, Massachusetts 02138, United States}
\author{Takatoshi Fujita}
\affiliation{Department of Chemistry and Chemical Biology, Harvard University, Cambridge, Massachusetts 02138, United States}
\author{Al\'an Aspuru-Guzik}
\email{Email: aspuru@chemistry.harvard.edu}
\affiliation{Department of Chemistry and Chemical Biology, Harvard University, Cambridge, Massachusetts 02138, United States}

%\contributor{Submitted to Proceedings of the National Academy of Sciences
%of the United States of America}

%% The \maketitle command is necessary to build the title page.
%\maketitle

%\begin{article}

\begin{abstract}
Phototrophic organisms such as plants, photosynthetic bacteria and algae use microscopic complexes of pigment molecules to absorb sunlight.
Within the light-harvesting complexes, which frequently have several  functional and structural subunits, the energy is transferred in the form of molecular excitations with very high efficiency. Green sulfur bacteria are considered to be amongst the most efficient light-harvesting organisms. Despite multiple experimental and theoretical studies of these bacteria the physical origin of the efficient and robust energy transfer in their light-harvesting complexes is not well understood.
To study excitation dynamics at the systems level we introduce an atomistic model that mimics a complete light-harvesting apparatus of green sulfur bacteria. The model contains approximately 4000 pigment molecules and comprises a double wall roll for the chlorosome, a baseplate and six Fenna-Matthews-Olson trimer complexes. We show that the fast relaxation within functional subunits combined with the transfer between collective excited states of pigments can result in robust energy funneling. Energy transfer is robust on the initial excitation conditions and temperature changes.  Moreover, the same mechanism describes the coexistence of multiple timescales of excitation dynamics frequently observed in ultrafast optical experiments. While our findings support the hypothesis of supertransfer, the model reveals energy transport through multiple channels on different length scales.
\end{abstract}

%\keywords{chlorosome | baseplate | light-harvesting complex
%| green sulfur bacteria | excitation energy transfer | photosynthesis}

%\abbreviations{BChl, bacteriochlorophyll; FMO, Fenna-Matthews-Olson; LHC, light-harvesting complex; EET, excitation energy transfer; IS, initial state}
%\keywords{Abbreviations: BChl, bacteriochlorophyll; FMO, Fenna-Matthews-Olson; LHC, light-harvesting complex; EET, excitation energy transfer; IS, initial state}

\maketitle
\section{Introduction}
\begin{figure*}[t]
\centerline{\includegraphics[width=2\columnwidth]{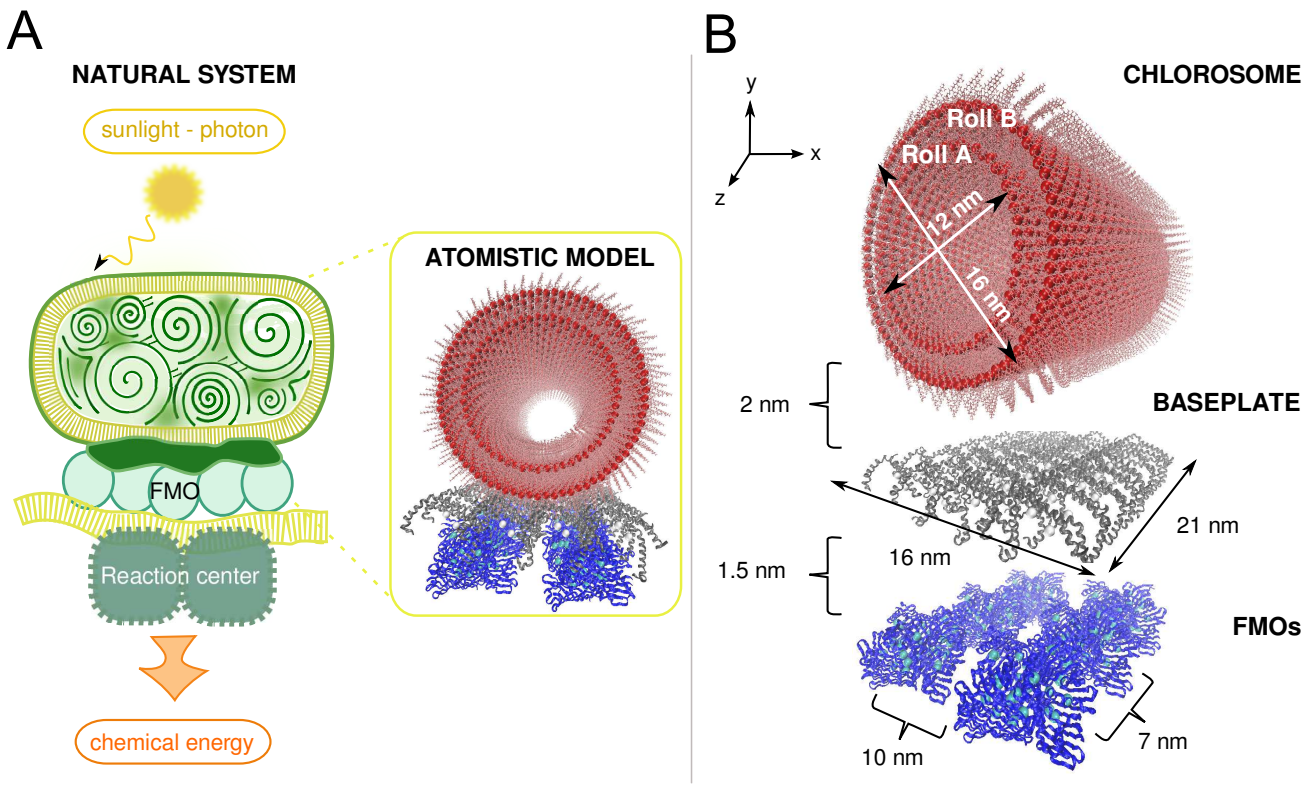}}
\caption{{\bf Photosynthetic apparatus.} {\bf A}, Cartoon of light-harvesting complex in green sulfur bacteria. The bacteria transform solar photons into chemical energy.
Sunlight absorbed by the chlorosome is transferred in the form of an exciton through
the baseplate and Fenna-Matthews-Olson (FMO) complexes subsequently to the reaction center. 
%The electro-chemical potential energy gradient is generated over a membrane and the energy is transformed as chemical energies finally.
A snapshot of the model structure is also shown. {\bf B}, Atomistic model with corresponding length scales. The atomistic model is composed of a double wall roll for the chlorosome (Roll A: 1620 (=60$\times$27) BChl \emph{c} sites and Roll B: 2160 (=80$\times$27) BChl \emph{c} sites), baseplate (64 BChl \emph{a} sites) and 6 FMO trimer complexes (144 (=24$\times$6) BChl \emph{a} sites).
\label{fig:cartoon}}
\end{figure*}
Photosynthetic bacteria are among the simplest organisms on Earth which use sunlight as their main energy source~\cite{blankenshipbook}. To collect solar energy these bacteria exploit light-harvesting complexes (LHC), aggregates of pigment molecules, which absorb photons and transfer the associated energy at the submicron scale.
The LHC in green sulfur bacteria contains large light absorbing antennae self-assembled in the so-called chlorosome~\cite{Oostergetel2010}.
These bacteria are obligate phototrophs -- they are required to use sunlight to support metabolic reactions \cite{Aquatic_book2005,Overmann_Prok7,Feng2010}. However, it has been observed that green sulfur bacteria can live in extremely low light conditions, even when receiving only a few hundred photons \emph{per  bacterium} per second \cite{Overmann1992,Beatty2005,Manske2005}. These facts have inspired many conjectures and discussions on the functional properties, energy conversion efficiency and robustness of LHC in green sulfur bacteria~\cite{Engel2007,Scholes2011,Borisov2012,Lloyd2010,Hoyer2010,Abasto2012,Kassal2013}. 

In order to address this controversy we introduce a model which includes atomistic structural detail of the green bacteria LHC and allows for the simulation of excitation energy transfer (EET) at the systems level. As a specific example, we consider the LHC of \textit{Chlorobium tepidum}.
We observe fast relaxation of excitations within the subunits of LHC owing to the large overlap between exciton states and strong interaction with environmental fluctuations. The transfer between subunits involves collective excited states of the pigment molecules and supports the hypothesis of supertransfer \cite{Lloyd2010,Abasto2012,Kassal2013}. The energy transport is robust to different initial excitation conditions, and changes in temperature. Finally, we show that the population of different parts of the LHC can be described using simple kinetic equations with time-dependent transfer rates characterizing intra-unit dynamics. This later model naturally explains the multiple timescales of EET reported in optical studies of green sulfur bacteria ~\cite{Savikhin1995,Pesencik1998,Prokhorenko2000,Psencik2003,Martiskainen2012a} and green non-sulfur bacteria~\cite{Fetisova1996,Shibata2007,Martiskainen2009}. 

Theoretical models have been applied mostly to single functional units of LHCs~\cite{Mukai1999,Jang2007a,Mohseni2008,Ishizaki2009,Strumpfer2009,Ritschel_Roden_Strunz_AspuruGuzik_Eisfeld2011,
Shim2012,Kreisbeck2012a,Kim2012} in order to understand the physical principles of energy transfer. Some of these studies also involved atomistic structures~\cite{Damjanovic2002,Olbrich2010,Shim2012,Kim2012}, which make the models computationally demanding. To the authors knowledge there are only a few atomistic studies of the complete light-harvesting systems of purple bacteria~\cite{Hu1997,Sener2007} but none for green sulfur bacteria.   
In addition to the large scale calculations the detailed analysis of excitation dynamics on the systems level~\cite{Hu1997,Sener2007,Martiskainen2009,Olaya-Castro2008,Ringsmuth2012} is complicated due to the lack of structural information. Thus, one usually needs to use macroscopic phenomenological models \cite{Dostal2012} or introduce additional constraints and approximations on the transport models~\cite{Cao2009,Caruso2009}. 
%Our work is, as mentioned already, the fully atomistic picture from chlorosome to cytoplasm of green photosynthetic bacteria.  

The LHC in green sulfur bacteria is composed of bacteriochlorophyll (BChl) pigment molecules. These monomers aggregate in several interconnected functional units, as shown in Fig.~\ref{fig:cartoon}{\bf A}. The main element of LHC is the chlorosome - an ellipsoidal shaped body of size ranging from tens to hundreds of nanometers \cite{Oostergetel2010}. The chlorosome is densely packed with BChl~$c$ pigments. Two other functional units - the baseplate~\cite{Pedersen2010} and the Fenna-Matthews-Olson (FMO) trimer complex~\cite{olson2004} are composed of BChl~$a$ pigments held together by a protein scaffolding. Energy in the form of molecular excitations (\emph{i.e.} exciton) is collected by the chlorosome and funneled through these antenna units to the reaction center where charge carriers are then generated.
The distance between the pigments in LHCs is sufficiently large such that the overlap of electronic wave functions can be neglected. In this case the energy transfer is mediated by the near field interaction between molecular electronic transitions, the F$\ddot{\mathrm{o}}$rster interaction~\cite{Forster1948,Maybook,Zimanyi2012}. If the interaction between several molecules is sufficiently strong as compared to the energy difference between their electronic transitions, the exciton states are delocalized over the group of pigments~\cite{Maybook,Zimanyi2012}. The preferential direction for energy transport is controlled by the frequencies of electronic transitions: the excitation goes to molecules or groups of molecules with lower excited state energy, while dissipating the energy difference to the environment.

\subsection{Molecular aggregate model}\label{sec:Model}
A single LHC of \textit{Chlorobium tepidum} contains 200--250 thousand BChl molecules \cite{Frigaard2003,Psencik2003,Oostergetel2010}.
Most of these molecules are found in the chlorosome. The model we have created is shown in Fig.~\ref{fig:cartoon}, it is composed of 3988 pigments and represents all the functional units of LHC in green sulfur bacteria, excluding the reaction center.

In our model (Fig.~\ref{fig:cartoon}{\bf B}) a double wall roll aggregate with diameter of about 16\,nm and length of about 21\,nm, represents the chlorosome. Several possible structural arrangements of BChls in the chlorosome have been investigated theoretically and experimentally~\cite{Holzwarth1994,Frese1997,Psencik2004,Linnanto2008,Ganapathy2009,Ganapathy2012,Tang2013}. Here we use the structure of Ref.~\cite{Ganapathy2009}, obtained from a triple mutant bacteria and characterized with nuclear magnetic resonance and cryo-electron microscopy. This structure is also supported by 2-dimensional polarization fluorescence microscopy experiments~\cite{Tian2011}. 

The microscopic structure of the baseplate has not yet been experimentally verified~\cite{Pedersen2010}. We construct a baseplate lattice as following. The unit cell consists of dimers of CsmA proteins~\cite{Pedersen2008} containing 2 BChl~\emph{a} molecules sandwiched between the hydrophobic regions and bound near the histidine. 
To establish a stable structure of the baseplate, classical molecular dynamics simulations were done. \footnote{The NAMD program package version 2.8~\cite{Phillips2005} was used. Force fields were  parameterized with a combination of Amber ff99SB for the protein~\cite{Hornak2006} and MMFF94 atomic charges for the BChl \emph{a}.}
The final structure complies with the periodicity and dimensions of the unit cell as seen in freeze frame fracture~\cite{blankenshipbook2}. %Moreover, the circular dichroism and fluorescence anisotropy spectra of the constructed baseplate agree with the measured spectra~\cite{Shibata2007,Pedersen2008}. 
The lattice model for the baseplate is described in the Appendix. 
Finally, for the FMO protein complexes we employ the structure resolved in Ref.~\cite{Tronrud2009}. 

The constructed model of the light-harvesting apparatus contains 95\,\% of BChl \emph{c} and 5\,\% of BChl \emph{a}, which is comparable to the stochiometry of the natural system (99\,\% and 1\,\%)~\cite{Psencik2003}. The estimated density of FMO complexes is about 1 FMO/50\,nm$^{2}$~\cite{Wen2009}.  Therefore, we distribute 6 FMO complexes under the baseplate which occupies about 300\,nm$^{2}$ (see Fig.~\ref{fig:cartoon}{\bf B}). This gives a pigment ratio of 2.3:1 (FMOs:baseplate), which is similar to the corresponding  stochiometry of \emph{Chlorobium tepidum} 2:1~\cite{Frigaard2003}. 

The distances between the chlorosome BChl \emph{c} aggregates and the baseplate is determined by the length of BChl \emph{c} esterifying alcohols. In the case of \textit{Chlorobium tepidum} it is about 2 nanometers~\cite{Martiskainen2009,Pedersen2010,Martiskainen2012a}. While the orientation of FMO relative to the baseplate has been verified experimentally~\cite{Wen2009}, the relative distance between these units is unknown. In our model we set it to be 1.5~nm, which is larger than the inter-pigment distance within FMO but smaller than the baseplate-chlorosome distance. This choice is based on the argument that the FMO complex is strongly linked to the  baseplate~\cite{SchmidtamBusch2011}. Minor variations of this distance do not affect the results. 

The frequencies of exciton transitions in LHCs are controlled by multiple factors. In the model it is equivalent to use the relative shifts (energy gap) of these transitions, which are relevant to the EET. These shifts can be calculated from the pigment-pigment couplings and the electronic excitations of single BChls, site energies, modified by the local environment~\cite{Adolphs2006}. While the couplings can straightforwardly be computed using a screened dipole-dipole model~\cite{Adolphs2006}, the calculation of site energies requires more complicated models or fitting to experimentally measured optical spectra.
Here, we set the frequency offset to be aligned with the lowest site energy of the FMO complex~\cite{Milder2010,SchmidtamBusch2011}. 

\begin{figure}[t]
\centerline{\includegraphics[width=1\columnwidth]{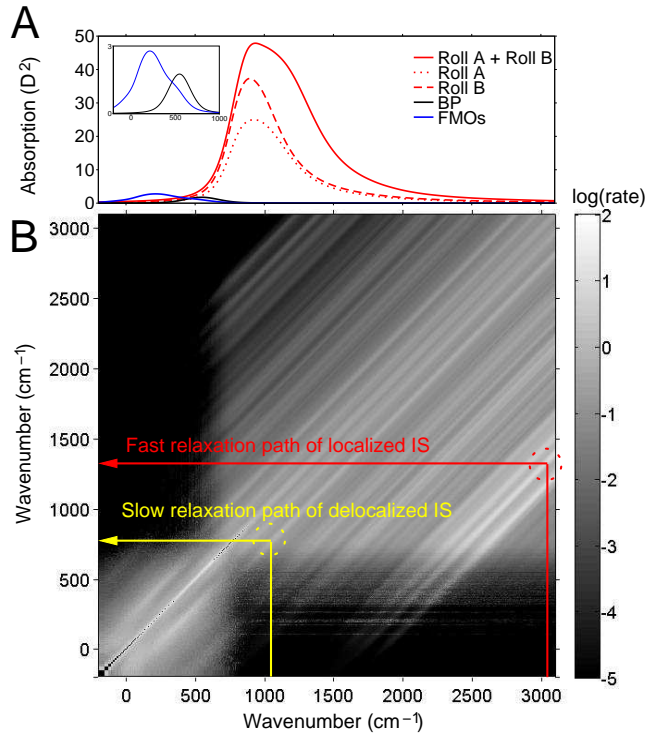}}
\caption{{\bf Calculated absorption spectra and exciton transfer rate matrix $\gamma_{MN}$.} {\bf A}, Calculated absorption spectra~\cite{Prokhorenko2003} by direct diagonalization of the system Hamiltonians of the antenna units in Fig.~\ref{fig:gammamatrix} are shown. The absorption spectra are calculated and drawn for the double wall roll (Roll A + Roll B), the single rolls (Roll A and Roll B), the baseplate and the 6 FMO complexes. The absorption spectrum of each antenna unit is obtained after taking 1000 ensemble average over the site energy fluctuations (static disorder). A Lorentzian line shape function with a full width at half maximum of 100 cm$^{-1}$ is convoluted, additionally, to take the homogeneous broadening into account. 
{\bf B}, Transfer rate matrix $\gamma_{MN}$ (cm$^{-1}$) at 300\,K is presented in a logarithmic scale.
$\gamma_{MN}$ indicates population transfer rate between exciton states $\vert M\rangle$ and $\vert N\rangle$. 
We set, here, the frequency offset to be aligned with the lowest site energy of the FMO complex. 
\label{fig:gammamatrix}}
\end{figure}

The absorption domains of the baseplate and FMO composed of BChl \emph{a} pigments are not clearly distinguishable. The absorption band of the baseplate covers the range 790--810\,nm. This range also includes the absorption band of the FMO complexes~\cite{Francke1997,Milder2010,Martiskainen2012a}. In fact, the absorption band of the baseplate significantly overlaps with that of the chlorosome~\cite{Martiskainen2012a}. In order to reproduce these spectra using the constructed model we define the site energy of BChl \emph{c} to be 2950\,cm$^{-1}$, which places the absorption maximum of the chlorosome of about 640\,cm$^{-1}$ above the absorption maximum of FMO complexes (see Fig.~\ref{fig:gammamatrix}{\bf A}). 
Our choice is based on the fluorescence maximum of the chlorosome (786 nm)~\cite{Martiskainen2012a}. We shift the lowest exciton state obtained after taking 1000 ensemble average over the site energy fluctuation (standard deviation: 500 cm$^{-1}$~\cite{Fujita2012}) at the fluorescence maximum.   
%As a result, the corresponding absorption maxima of BChl \emph{a} and \emph{c} domains are found at 804\,nm and 759\,nm, respectively, 
12225\,cm$^{-1}$ (818\,nm) is used as the offset energy value. 
We assign the site energy of the baseplate as 550 cm$^{-1}$, which places the absorption maximum of the baseplate approximately in the middle of the absorption maxima of the FMO complexes and the chlorosome. 
The resulting absorption spectrum of the baseplate is shown in Fig.~\ref{fig:gammamatrix}{\bf A}.

\subsection{Exciton transfer model}
The exciton transfer is modelled with a quantum master equation approach, which includes the coherent, dephasing and relaxation  processes, for the open quantum dynamics~\cite{Louisellbook,Breuerbook,Adolphs2006,Rebentrost2009,Rebentrost2009b}.
We solve the quantum master equation to obtain the spacial distribution of the exciton.   

In our model, the system-bath Hamiltonian of the light-harvesting apparatus is composed of three parts:
the system consists of the local excitations of bacteriochlorophylls (BChls) and the point dipole interactions between them, described using a tight-binding Hamiltonian. Then, the system (BChls) is coupled linearly to the bath (proteins). The bath Hamiltonian consists of a sum of multidimensional quantum harmonic oscillators (see \emph{e.g.} Ref.~\cite{Rebentrost2009b} and the Appendix). 

Within the secular approximation and in the Markov limit (\emph{i.e.} secular Redfield), the equations of motion of the reduced density operator $\hat{\rho}_{\mathrm{S}}(t)$ in the exciton basis,  
%(\emph{cf.} Ref.~\cite{Kubota2008}) is written in a generalized Lindblad quantum master equation form~\cite{Breuerbook,Kondov2003}.
the population and coherence transfer are decoupled~\cite{Ishizaki2009c}. 
The equations of motion is given in the Appendix.  
%After ignoring pure dephasing terms (see \emph{e.g.} Ref.~\cite{Kreisbeck2012a} for the discussion) and degeneracy of the exciton states (numerical degeneracy within 1\,cm$^{-1}$)~\cite{Louisellbook}, the final quantum master equation is given in a standard Lindblad form~\cite{Breuerbook}. 

The resulting quantum master equation includes a term $\gamma_{MN}$, which is the exciton transition rate between the corresponding exciton states $\vert M\rangle$ and $\vert N\rangle$. $\gamma_{MN}$ is calculated
with the exciton eigenvectors and spectral density (exciton-phonon coupling strength) at the transition energy (see Ref.~\cite{Valleau2012b,Rebentrost2009b} for the definition and also the Appendix for the expressions). It is shown in Fig.~\ref{fig:gammamatrix}{\bf B} as a matrix, for the EET dynamics at 300\,K. 

%We describe the exciton dynamics 
%with the secular Redfield quantum master equation,
%which includes coherent, dephasing and relaxation processes~\cite{Louisellbook,Breuerbook,Adolphs2006,Rebentrost2009,Rebentrost2009b}.
%\cjoon{The concrete model is described in the Appendix.} 
The validity of the Redfield method for the EET in natural light-harvesting structures had been discussed by many authors, see \emph{e.g.} Refs.~\cite{Zhang1998,Pullerits2000,Yang2002,Novoderezhkin2003,Novoderezhkin2004,Ishizaki2009c} and the references cited therein. 
%Yang and Fleming~\cite{Yang2002} claimed, for example, that the validity of the Redfield equation relies on the frequency domain of the phonon spectral density rather than the relatively weak exciton-phonon coupling strengths with respect to the pigment-pigment couplings. 
When the energy gap between the exciton states is small, the Redfield model with a broad spectral density can be applicable~\cite{Yang2002}. 
Our molecular aggregate model in Fig.~\ref{fig:cartoon} and the corresponding spectral densities~\cite{Shim2012,Fujita2012} satisfy this condition: the absorption spectra of the antenna units overlap each other significantly, which implies the exciton states in this energy domain are delocalized over the two antenna units. The antenna units are coupled weakly (16--17 cm$^{-1}$).   
The spectral densities and the density of states are given in the Appendix. 
 
Novoderezhkin and \emph{et al.}~\cite{Novoderezhkin2003} proposed to compensate the underestimation of the transfer rate between exciton states with large energy gaps by increasing the spectral density in the high frequency region. 
Therefore, we note here the exciton transition rate, which involves the exciton transfer with a large energy gap, could be underestimated because the Redfield model can only account for the single phonon process. Multiphonon processes could occur in the internal exciton dynamics of the antenna units due to its broad exciton bands (see Fig.~\ref{fig:gammamatrix}{\bf A}). The internal exciton dynamics of the chlorosome is, however, much faster than the exciton transfer between the antenna units. Thus, the Redfield model should give a reasonable results (timescales) qualitatively for the exciton funneling process of the photosynthetic apparatus.
For more accurate models, one would consider other methods such as the modified Redfield approach~\cite{Zhang1998,Yang2002,Novoderezhkin2004}, hierarchical equations of motion~\cite{Ishizaki2009b,Strumpfer2009,Kreisbeck2011,Zhu2011}, iterative linearized density matrix dynamics~\cite{Huo2011}, non-Markovian quantum state diffusion~\cite{Ritschel2011,Ritschel_Roden_Strunz_AspuruGuzik_Eisfeld2011}, variational master equation~\cite{Mccutcheon2011}, path integral Monte Carlo~\cite{Muhlbacher2012},  and see the references cited in the review~\cite{Pachon2012} of the methodologies in EET. However, most of these sophisticated methods, are not applicable to our large system because they are numerically too demanding.

The effects of slow fluctuations in the site energies (static disorder), which are responsible for the inhomogeneous broadening, are incorporated. We use 100 $\rm{cm}^{-1}$ for the Gaussian fluctuations in FMO and the baseplate, and 500 $\rm{cm}^{-1}$ for the roll~\cite{Milder2010,Fujita2012}.
All results are obtained from 1000 ensemble averages for the static disorder, unless otherwise mentioned.

The system Hamiltonian of FMO trimer complexes is taken from Ref.~\cite{SchmidtamBusch2011}. The spectral density from our previous work~\cite{Shim2012} is used: where molecular dynamics and time-dependent density functional theory calculations were used for obtaining it. A harmonic prefactor was used for the spectral density~\cite{Valleau2012b}.  The structure of the double wall roll is obtained based on Ref.~\cite{Ganapathy2009}. The structure was optimized with molecular dynamics simulation and a spectral density was obtained by time-dependent density functional theory  calculations~\cite{Fujita2012,Valleau2012b}. 

Instead of computing the spectral density of the baseplate, which is composed of BChl \emph{a}, we use the spectral density of  FMO~\cite{Shim2012}. This approximation is justified because we expect the vibrational structure to be similar to FMO's, which is surrounded by a protein environment (\emph{cf.} chlorosome) and is also composed of BChl \emph{a}. 
%\cjoon{The spectral density of the baseplate obtained by semi-empirical / molecular dynamics calculations is similar to the spectral density of the FMO complex in the low frequency domain (500 cm$^{-1}$), which is mainly responsible for the exciton transfer between the chlorosome and the baseplate. in the Appendix, we present the spectral densities obtained by semi-empirical / molecular dynamics calculations. However, this spectral density is not used for the baseplate because we want to use the spectral densities from the same method (\emph{i.e.} time-dependent density functional theory).}

%However, we expect a larger reorganization energy for the baseplate because the Stokes shift of the baseplate is about 300\,cm$^{-1}$~\cite{Martiskainen2012a} while the Stokes shift of the FMO trimer complex is small~\cite{Savikhin1997}.

To this end, we define the mean exciton energy to quantify the energy dissipation
from the system to the bath during the energy funneling process,
\begin{align}
\mathrm{MEE}(t)=\mathcal{E}\left(\mathrm{Tr}_{\mathrm{S}}\left(\hat{H}_{\mathrm{S}}\hat{\rho}
_{\mathrm{S}}(t)\right)\right)
,\label{eq:mee}
\end{align}
where $\hat{H}_{\mathrm{S}}$ is the system Hamiltonian and $\mathrm{Tr}_{\mathrm{S}}$
is the trace over the system degrees of freedom. 
$\mathcal{E}$ is the ensemble average over the static disorder. 

Additionally, we introduce the cooperativity, which is used to quantify the enhancement of transition dipole moment by coherence. Cooperativity$(t)$ can be interpreted as excitation delocalization, as following
\begin{align}
&\mathrm{Cooperativity}(t)= \nonumber \\  
&\frac{1}{\vert\mu\vert^{2}}
\mathcal{E}
\left(\sum_{\alpha=x,y,z}
\sum_{m,n \in \mathrm{Domain}}\mu_{n,\alpha}\mu_{m,\alpha}
\langle n \vert\hat{\rho}_{\mathrm{S}}(t)\vert m \rangle
\right),
\label{eq:coop}
\end{align}
where $\mu_{n}$ is the transition dipole moment vector of site $n$ and
a normalization factor $\vert\mu\vert^{2}=30$ D$^{2}$, 
which is the absolute square of the transition dipole moment 
of a single pigment, is used. All pigments have the same magnitude of 
the transition dipole moment in our model (Fig.~\ref{fig:cartoon}{\bf B}). $\vert m \rangle$ and $\vert n \rangle$ are the site basis states. 
The summation is over the domain of interest.

\section{Excitation energy funneling} \label{sec:EET}
\begin{figure}[t]
\centerline{\includegraphics[width=1\columnwidth]{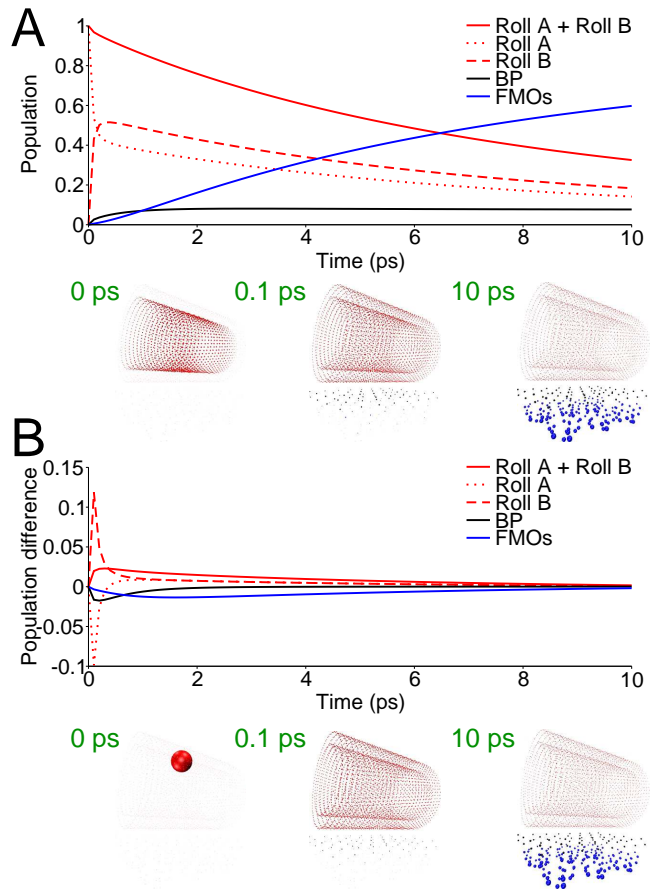}}
%\centerline{\includegraphics[width=1\columnwidth]{figures/FIG4B.eps}}
\caption{{\bf Exciton population dynamics with a delocalized and a localized initial state at 300\,K.}
{\bf A}, The initial state (see the snapshot at 0\,ps) is the brightest state of Roll A.
{\bf B}, The initial state (see the snapshot at 0\,ps) is a localized state, \emph{i.e.} a site on the top and in the middle of the Roll A.
The population difference with respect to the population in {\bf A} are plotted. 
The locations of magnesium (Mg) in the BChls represent the locations of exciton sites and the sizes of the spheres are proportional to the populations of the corresponding sites. The populations of the rolls, the baseplate and the FMOs are designated red, black and blue, respectively. 
\label{fig:popen}}
\end{figure}

To fully characterize the exciton transfer process of the photosynthetic apparatus model in Fig.~\ref{fig:cartoon}, one needs to study the exciton dynamics for all possible initial (exciton) states within an ensemble at a finite temperature. 
For instance, the initial state prepared by a coherent light source (laser) could be considered as a single exciton state~\cite{Brumer2012}.
%Instead of studying all possible initial exciton states, 
As an example, we perform exciton dynamics simulations for two cases of initial excitation at 300\,K to see how the initial condition affects the EET dynamics.
One is the brightest exciton state of the system Hamiltonian of Roll A, which is delocalized over Roll A (see the snapshot of Fig.~\ref{fig:popen}{\bf A} at 0\,ps) and has energy 1018\,cm$^{-1}$.
The other initial condition to be considered is a localized initial state (IS). In particular, a single site located on top and in the middle of Roll A is selected for the localized IS having energy 3022\,cm$^{-1}$ (see the snapshot of Fig.~\ref{fig:popen}{\bf B} at 0\,ps).

Comparing the absorption spectra of Roll A and Roll B in Fig.~\ref{fig:gammamatrix}{\bf A}, one can see the peak maximum of Roll B is red-shifted from the peak maximum of Roll A, thus there is an exciton energy gradient between the layers.
As the radius of the roll increases (contrast A and B), the peak maximum shifts to the red~\cite{Fetisova1996,Prokhorenko2003,Linnanto2008}. This occurs because the roll curvature changes and this induces stronger dipole-dipole interactions between neighboring pigments. This energy gradient is favorable for the exciton energy funneling because EET from the outermost layer to the baseplate is important.
Our choice of the initial states on the Roll A is based on this argument. 

There are two important factors in determining the exciton transfer between the antenna units. These are the energy resonance condition and the electronic coupling between the energy levels of the antenna units~\cite{Jang2004}. The former is the necessary condition for the EET between the units and the later determines how fast EET should be.
Fig.~\ref{fig:gammamatrix} shows the delocalized IS is close to the energy levels of the baseplate and large multichromophoric excitonic coupling strengths to the baseplate exciton states. 
In contrast, the localized IS is far from the energy resonance level to the baseplate and the excitonic coupling strength is small.

Fig.~\ref{fig:popen} summarizes the resulting exciton dynamics at 300\,K.
Figs.~\ref{fig:popen}{\bf A} and~\ref{fig:popen}{\bf B} show the population dynamics using the delocalized IS and the localized IS respectively and up to 10\,ps. Our choice of the time interval (10\,ps) of the EET simulation is based on the timescales of the EET of \emph{Chlorobium tepidum} in Ref.~\cite{Martiskainen2012a}.
Snapshots of the site populations at 0\,ps, 0.1\,ps and 10\,ps are shown below the population plots.
The exciton population distributions of individual antenna units at 10\,ps are almost identical regardless of the initial conditions. 
For example, the total exciton  population on FMOs is approximately 60\,\% for the two initial conditions. 

In the rest of this section, we provide more detailed discussion of severe aspects of the exciton transfer. 
First, the exciton population dynamics 
of the two initial conditions are compared. Then, the multichromophoric 
effect is discussed for the exciton dynamics. The temperature dependence of the exciton dynamics comes afterwards. Lastly, the exciton dynamics is described in terms of the population kinetic model.   
\begin{figure}[t]
\centerline{\includegraphics[width=1\columnwidth]{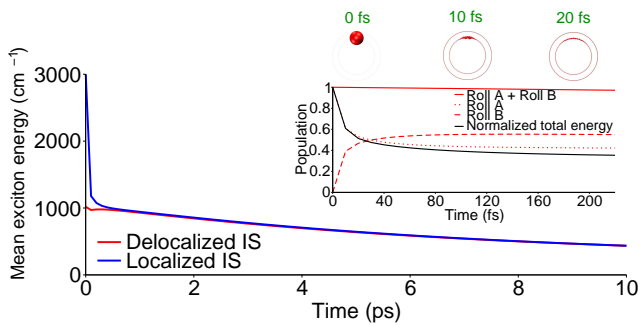}}
\caption{{\bf Mean exciton energy with a delocalized and a localized initial state at 300\,K.}
Mean exciton energy (Eq.~\ref{eq:mee}) with the two different initial excitations, which correspond to those in Fig.~\ref{fig:popen}{\bf A} and~\ref{fig:popen}{\bf B}. The short time dynamics for the first 200\,fs of Fig.~\ref{fig:popen}{\bf B} is given in the inset with the corresponding population snapshots. The populations in the snapshot which are projected to the long axis of the rolls. The locations of magnesium (Mg) atoms in the BChls represent the locations of exciton sites and the sizes of the spheres are proportional to the populations of the corresponding sites.
\label{fig:mee}
}
\end{figure}

\subsection{Exciton population dynamics}
The EET dynamics of the delocalized IS and the localized IS become similar within 1\,ps (Fig.~\ref{fig:popen}).
The short time dynamics ($<$ 200\,fs), however, are sufficiently different. Fig.~\ref{fig:popen}{\bf A} shows a fast initial population decay for the roll comparing to that of the localized IS in Fig.~\ref{fig:popen}{\bf B}. Characteristic time constants in Table~\ref{tab:timeconst} are extracted by the exponential fitting of the exciton populations of the roll such that amplitudes are summed to be 100\,\%.
By comparing the time constants for the roll in Table~\ref{tab:timeconst}, we see that Set I (delocalized IS) has a fast sub-100-fs component while Set III (localized IS) does not. However, $\tau_{1}$ in Set I accounts for only 3\,\% of the 10\,ps exciton dynamics.
In the case of using the delocalized IS, the single exciton starts to migrate from the roll to the baseplate already at the very beginning ($<$ 100\,fs). 
This occurs because the frequency of the delocalized IS (1018\,cm$^{-1}$) is close to the baseplate absorption region (see Fig.~\ref{fig:gammamatrix}{\bf A}) and has a large collective transition dipole moment.\footnote{Another delocalized IS, which is the brightest exciton state of Roll B, 
shows the similar short time dynamics (see Set II and the Appendix for the corresponding time constants and the exciton dynamics, respectively).} 
In contrast, the localized IS (3022\,cm$^{-1}$) is far from the energy resonant region and has a comparably weak transition dipole moment. 

Equilibration in the roll is achieved within 100\,fs for the localized IS dynamics and almost no exciton population is transferred to the baseplate in this short time. This can be seen in the inset figure of Fig.~\ref{fig:mee}.
The inset in Fig.~\ref{fig:mee} shows the diffusion process in the roll with the localized IS for the first 200\,fs. Snapshots of the roll populations at 0, 10 and 20\,fs are placed above the inset plot. In this plot, one can see how the single exciton diffuses within and between the layers. The black solid line in the inset figure is the mean exciton energy (Eq.~\ref{eq:mee}), which is normalized to the initial energy (3022\,cm$^{-1}$).
Interestingly, the curve is similar to the population dynamics of Roll A. From this we can conclude that the population transfer from Roll A to Roll B is the main energy relaxation channel and the slight difference of the two curves indicates the effect of population redistribution within the single layers. Thus the energy dissipation due to exciton-phonon coupling mainly causes exciton transfer between the layers in this initial short time period.
The mean exciton energy of the total system (Roll+baseplate+FMOs) is given in Fig.~\ref{fig:mee} for two different initial excitations.
The solid blue line and the solid red line correspond to the exciton dynamics of the two different initial conditions, respectively.

As mentioned above, the initial energy of the delocalized state is already close to the baseplate bright state energy domain (see Fig.~\ref{fig:gammamatrix}), while that of the localized state is higher (3022\,cm$^{-1}$).
In the localized IS case, the excess energy (about $2000$\,cm$^{-1}$) should be released to the environment in order for resonant energy transfer to the baseplate to occur.
In spite of the high initial energy of the localized IS, which is far from the energy resonance domain, the exciton population of each unit at 10\,ps is similar to that of the delocalized IS case (see Fig.~\ref{fig:popen}{\bf A}). This is possible because a rapid energy relaxation channel (Fig.~\ref{fig:gammamatrix}{\bf B}) is available for the dynamics of Set III. The blue line in Fig.~\ref{fig:mee} shows a rapid energy drop within 100\,fs.
Then, within 1\,ps, the total energy approaches the energy of the delocalized IS. The population snapshots at 100\,fs indicate that population distributions are quite similar. Also, the population on the roll in the snapshots of Fig.~\ref{fig:popen}{\bf B} at that time indicates that, by 100\,fs, the system population is mostly delocalized over the roll.
The mean exciton energy obtained from the exciton dynamics with the delocalized IS and the localized IS become similar within 500\,fs. 
The rapid relaxation within the roll results in robust energy transfer from the roll to the FMOs in the long time limit in our model study.

Microscopically, the energy dissipation dynamics is determined by thermal excitations and relaxation among exciton levels. The energy dissipation rate, in this model, depends on the spectral density, a quantity which indicates how strongly exciton states are coupled to the thermal bath, the probability distribution of the exciton states and temperature.

In Fig.~\ref{fig:gammamatrix}{\bf B}, we show the exciton transfer matrix ($\gamma_{MN}$) at 300\,K in logarithmic scale (log(cm$^{-1}$)).
We indicate the fast energy dissipation path for the localized IS with a red arrow. The strong white diagonal band corresponds to the strong exciton-phonon coupling at 1600--2000\,cm$^{-1}$~\cite{Roden_Schulz_Eisfeld_Briggs_2009} (see the spectral densities in the Appendix), which leads to the rapid energy dissipation of the localized IS within the roll.  
We note here that this fast relaxation occurs only between the exciton states in the same antenna units not between the exciton states of different antenna units.
 
Damjanovi$\acute{\mathrm{c}}$ \emph{et al.}~\cite{Damjanovic2002} 
suggested that a weakly bound polaron can be formed 
in BChl aggregates due to the interaction of excitons with intramolecular vibrational 
mode at about 1670\,cm$^{-1}$. Their results were based on studies of LHC in purple bacteria. We do expect that the polaron couplings can renormalize energy levels 
and the mobility of the exciton energy is reduced~\cite{Damjanovic2002}. 
This should be, however, weaker in the chlorosome where BChls are densely packed and 
the pigment-pigment interaction is, accordingly, stronger than that of LHC in purple bacteria. 

The exciton dynamics in the FMOs is conditioned mainly by the population of the baseplate because direct population transfer from the roll to FMOs is negligible (see the Appendix). 
%Therefore, the short time dynamics of FMOs are different from each other in Figs.~\ref{fig:popen}{\bf A} and~\ref{fig:popen}{\bf B}.
%Exciton populations of FMOs increase linearly and parabolically in the short time, respectively, in Figs.~\ref{fig:popen}{\bf A} and~\ref{fig:popen}{\bf B}. However, after 3\,ps, the population dynamics converge to each other.

\begin{table}[b] %~~~~~~~~~~~~~~~~~~~~~~~~~~~~~~~~~~~~~~~~~~~~~~~~~~~~~~~~~~~~~~

\begin{centering}

\caption{{\bf Time constants of the exciton dynamics of the chlorosome roll.} The values are obtained by the exponential fittings ($A_{1}\exp(-t/\tau_{1})+A_{2}\exp(-t/\tau_{2})+A_{3}\exp(-t/\tau_{3})$) of the exciton population dynamics for each antenna unit. The amplitudes ($A_{1}$, $A_{2}$ and $A_{3}$) are summed to be 100\,\%.\label{tab:timeconst}}

%\begin{tabular}{ccrrr}
%\toprule
%Set& $\tau_{1}(\mathrm{fs})$ ($A_{1}(\%)$) & $\tau_{2}(\mathrm{ps})$ ($A_{2}(\%)$)& $\tau_{3}(\mathrm{ps})$ ($A_{3}(\%)$)\\
%\midrule
%I\tablenote{Corresponds to the exciton dynamics of Fig.~\ref{fig:popen}{\bf A}} & 91.2 (6) & 1.9 (22) & 8.7 (72)\\
%II\tablenote{Brightest delocalized initial state of Roll B is used.}& 109.4 (8) & 2.4 (26) & 9.5 (66)\\
%
%III \tablenote{Corresponds to the exciton dynamics of Fig.~\ref{fig:popen}{\bf B}}& - & 3.7 (57) & 13.9 (43)\\
%
%IV\tablenote{The baseplate is not included in the exciton dynamics.}& - & 10.7 (5) & 1500.0 (95)\\
%
%Ref.~\cite{Martiskainen2012a}\tablenote{Anisotropic decay of \emph{Chlorobium tepidum} at 807\,nm} & - & 1.1(42) & 12.1(58)\\
%\bottomrule
%\end{tabular}
\begin{tabular}{ccrrr}
\hline
Set& $\tau_{1}(\mathrm{ps})$ ($A_{1}(\%)$) & $\tau_{2}(\mathrm{ps})$ ($A_{2}(\%)$)& $\tau_{3}(\mathrm{ps})$ ($A_{3}(\%)$)\\
\hline
I\tablenote{Corresponds to the exciton dynamics of the delocalized IS} & 0.081 (3) & 4.5 (36) & 12.9 (61)\\
II\tablenote{Brightest delocalized initial state of Roll B is used. 
The exciton dynamics is given in the Appendix.}& 0.060 (3) & 4.7 (39) & 13.1 (58)\\

III \tablenote{Corresponds to the exciton dynamics of the localized IS}& - & 3.7 (25) & 11.3 (75)\\

%IV\tablenote{The baseplate is not included in the exciton dynamics.}& - & 10.7 (5) & 1500.0 (95)\\
%IV \tablenote{Brightest delocalized initial state of Roll A is used at 150 K.}& 0.036 (1) & 5.4 (59) & 13.0 (4)\\
%V \tablenote{Brightest delocalized initial state of Roll A is used at 77 K.}& 0.078 (1) & 7.7 (89) & 28.9 (10)\\

Ref.~\cite{Martiskainen2012a}\tablenote{Anisotropic decay of \emph{Chlorobium tepidum} at 807\,nm} & - & 1.1(42) & 12.1(58)\\
\hline
\end{tabular}
\par\end{centering}

\end{table} %~~~~~~~~~~~~~~~~~~~~~~~~~~~~~~~~~~~~~~~~~~~~~~~~~~~~~~~~~~~~~~~~~~

\subsection{Cooperativity of the excitonic states}
In multichromophoric systems, coherent coupling between donor molecules can lead to a large collective transition dipole moment. This enhances  the energy transfer from the donor to acceptor groups as compared to incoherent hopping between individual molecules~\cite{Causgrove1992,Lloyd2010,Abasto2012,Kassal2013}.

\begin{figure}[t]
\centerline{\includegraphics[width=1\columnwidth]{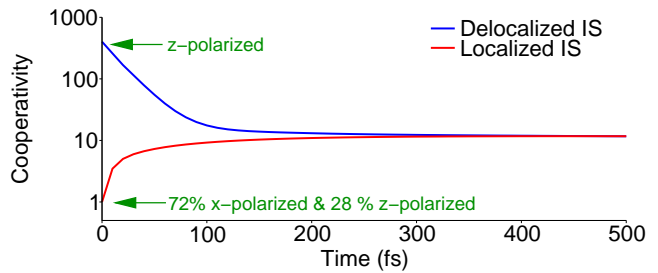}}
\caption{{\bf Time dependent cooperativity of the chlorosome at 300\,K.}
The cooperativity (Eq.~\ref{eq:coop}), dimensionless normalized collective transition dipole moment, is given in logarithmic scale.
During the exciton dynamics in Fig.~\ref{fig:popen} with the two initial states, the cooperativities are calculated for the chlorosome. 
%Two initial states are tested i) a delocalized IS for the brightest state of Roll A (blue) ii) a localized IS on a single site on top and in the middle of Roll A (red) and lastly iii) a delocalized IS on a single ring in the middle which has an equal amplitude and phase over the ring (black). 
\label{fig:mtdm}
}
\end{figure}

In Fig.~\ref{fig:mtdm}, we show the cooperativity (Eq.~\ref{eq:coop})  computed for first 500\,fs. The cooperativity is calculated for the two different initial excitation conditions corresponding to the dynamics in Fig.~\ref{fig:popen}. 
% and lastly iii) a delocalized IS on a single ring in the middle which has an equal amplitude and phase over the ring (black).

The delocalized IS is z-polarized (along the length of the roll) and initially has a cooperativity of 402 (out of 1620 pigments in the Roll A). This strong collective oscillator strength can induce rapid supertransfer~\cite{Lloyd2010,Abasto2012,Kassal2013}. 
%The delocalized IS (ring) is also z-polarized because the selected initial state has the equal phase over all components. The initial cooperativity of the delocalized IS (ring) is 16. 
The localized IS, which is 72\,\% x- and 28\,\% z-polarized, has an initial cooperativity value of unity. This difference in cooperativity at varying initial condition is one of the reasons why a fast decay component is found for the delocalized IS case only. 
%However, the cooperativity of the delocalized IS drops quickly within 500\,fs.

Regardless of the initial conditions, within 500\,fs, all cooperativity values converge to a similar value ($\sim$12 out of 3780 pigments in the Roll A and B), which is still larger than 1, and the effective transition dipole moment becomes about 30\,\% x-, 30\,\% y- and 40\,\% z-polarized. This is a favorable situation, for our photosynthetic apparatus model, as y-polarization (normal direction to the baseplate) is useful to funnel energy towards the baseplate.
These results may indicate a multichromophoric effect~\cite{Jang2004}; \emph{i.e.} the effective dipole moment of the delocalized exciton state is enhanced by symmetry (see also Ref.~\cite{Strumpfer2012} for the discussion on the coherence and EET rate). 
%, thus the exciton transfer coupling strength of exciton states of roll and the baseplate becomes large enough for rapid transfer even though the chromophores are far apart ($>$ 2.0\,nm) (see also Ref.~\cite{Strumpfer2012} for the discussion on the coherence and EET rate).

\subsection{Temperature dependence of the energy funneling}
\begin{figure}[t]
\centerline{\includegraphics[width=1\columnwidth]{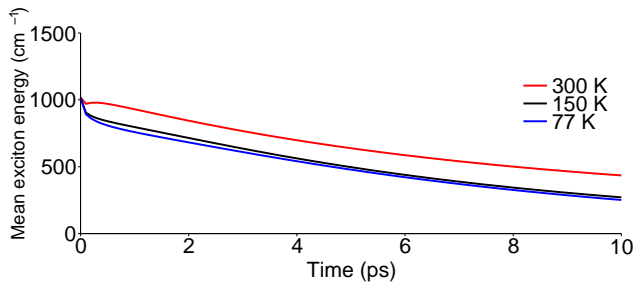}}
\caption{{\bf Temperature dependence of the mean exciton energy.}
The mean exciton energy of the system (Eq.~\ref{eq:mee}) is plotted at 300\,K, 150\,K and 77\,K.
The delocalized exciton initial state in Fig.~\ref{fig:popen}{\bf A} is used.
\label{fig:temperature}
}
\end{figure}

In the previous subsections, we showed that the exciton energy funneling process is robust to variations in initial excitation conditions due to the fast internal exciton dynamics of the roll. We now investigate the temperature effect by simulating the exciton population dynamics with the delocalized IS initial condition, \emph{i.e.} the brightest state of Roll A, at 150\,K and 77\,K in Fig.~\ref{fig:temperature}.

The mean exciton energy at room temperature (300\,K) in Fig.~\ref{fig:popen}{\bf A} is only slightly different from the curves at 150\,K and 77\,K in Fig.~\ref{fig:temperature}. 
The corresponding exciton population dynamics are given in the Appendix.  
This indicates that exciton transfer is robust within this temperature range.
The robust energy transfer within the temperature range is due to the fast internal exciton dynamics of the roll. The thermal excitation within the temperature range does not lift the exciton far from the energy resonance domain between the roll and the baseplate.

Thermal excitation in the temperature range (77, 150 and 300\,K) can provide various channels towards the neighboring exciton states for the relaxation process (see Fig.~\ref{fig:gammamatrix}{\bf B}). Thermal excitation can also induce back transfer from  the baseplate to  the rolls~\cite{Causgrove1992} but it reduces the possibility of being trapped in dark states.

\subsection{Population kinetics}
So far, we have shown that regardless of initial conditions and temperature about 60--70\,\% of the exciton can be transferred to the FMOs within 10\,ps. This robustness to the choice of initial conditions implies that internal dynamics within the roll is faster than energy transfer between the antenna units.
We now proceed to examine the population dynamics by using a simple first order kinetic model (more sophisticated kinetic models in the EET of the LH complex networks can be found in \emph{e.g.} Refs.~\cite{Yang2003,Caruso2009,Cao2009,Valkunas2011}):

\begin{align}
\frac{\mathrm{d}}{\mathrm{d}t}
\left(
\begin{smallmatrix}
[\mathrm{R}](t)\\
[\mathrm{BP}](t)\\
[\mathrm{FMO}](t)
\end{smallmatrix}
\right)
=
\left(
\begin{smallmatrix}
-k_{\mathrm{RBP}}(t) & 0 & 0\\
k_{\mathrm{RBP}}(t) & -k_{\mathrm{BPFMO}}(t) & 0\\
0 & k_{\mathrm{BPFMO}}(t) & 0
\end{smallmatrix}
\right)
\left(
\begin{smallmatrix}
[\mathrm{R}](t)\\
[\mathrm{BP}](t)\\
[\mathrm{FMO}](t)
\end{smallmatrix}
\right)
\end{align}
where $[\cdot](t)$ denotes the population of each antenna unit, and [R] and [BP] are the population of the full roll (Roll A + Roll B) and the baseplate respectively. 
$k_{\mathrm{RBP}}(t)$ is the exciton transfer rate from the roll to the baseplate and
$k_{\mathrm{BPFMO}}(t)$ is the one from the baseplate to the FMOs.
The population transfer between the antenna units is characterized by time-dependent rate constants $k(t)$. Note that the internal dynamics within the antenna units, such as relaxation and thermal excitation among the exciton states, is incorporated into the time dependence of $k(t)$.   
$k(t)$, physically, corresponds to the multichromophoric F$\ddot{\mathrm{o}}$rster resonance energy transfer rate~\cite{Jang2004}, because it quantifies energy transfer between the donor group (exciton states) of the roll and the acceptor group of the baseplate. The enhancement of energy transfer due to coherence (Fig.~\ref{fig:mtdm}) between donor molecules is also referred as to supertransfer~\cite{Lloyd2010,Abasto2012,Kassal2013}. 

\begin{figure}[t]
\centerline{\includegraphics[width=1\columnwidth]{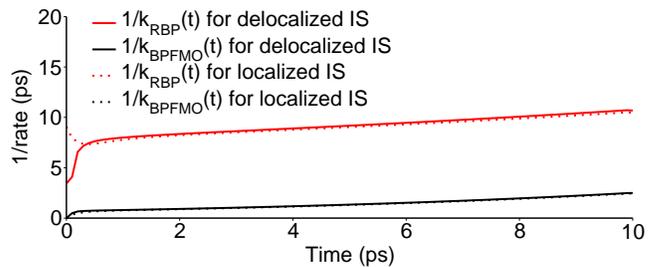}}
\caption{{\bf Time dependent reciprocal rate.} Time-dependent exciton transfer rates are given as the corresponding time constants, the reciprocal rate, for the exciton dynamics of the delocalized IS and the localized IS in Fig.~\ref{fig:popen}.
$k_{\mathrm{RBP}}(t)$: exciton transfer rate from the roll to the baseplate.
$k_{\mathrm{BPFMO}}(t)$: exciton transfer rate from the baseplate to the FMOs.
\label{fig:rate}}
\end{figure}

%To test if there is direct exciton transfer from the roll to the FMO, 
%we simulate the exciton dynamics without the baseplate in the energy funneling process (see the Appendix). 
%\cjoon{There is almost no direct population transfer from the roll to the FMO complexes. Only 1\,\% of the exciton is transferred to the FMO complexes within 10 ps without the baseplate.} 
%Set IV in Table~\ref{tab:timeconst} shows the result, the population dynamics of the roll
%is composed of 10\,ps and 1.5\,ns timescales with 5\,\% and 95\,\% amplitudes respectively.
%The exciton transfer from the roll to FMOs is extremely slow such that only 1\,\% is transferred within 10\,ps. 
The direct exciton transfer from the roll to the FMO complexes is 
virtually negligible within the time interval of the EET dynamics 10\,ps (see the Appendix). 
In this kinetic model, thus, we assume there is no population transfer from low to high energy units and no direct transfer from the roll to FMOs. 
The kinetic models are fitted to the exciton populations in Fig.~\ref{fig:popen} using least squares. The resulting time-dependent
population transfer rates are shown in Fig.~\ref{fig:rate} for the exciton dynamics of Fig.~\ref{fig:popen}, using both initial conditions (the delocalized IS and the localized IS). The initial and final values of the reciprocal rates of the (chlorosome) roll $1/k_{\mathrm{RBP}}(t)$ have similar values to $\tau_{2}$ and $\tau_{3}$ of Sets I and III in Table~\ref{tab:timeconst}. Within 500\,fs, $k_{\mathrm{RBP}}(t)$ for the delocalized IS drops rapidly to a slower rate, with a similar timescale to the equilibrium time of the cooperativity (Eq.~\ref{eq:coop}), see solid blue line in Fig.~\ref{fig:mtdm}. However we see that $k_{\mathrm{RBP}}(t)$ for the localized IS does not show this rapid drop.
Regardless of the initial conditions, the rate constants become similar to each other within 500\,fs.
As could be expected, $k_{\mathrm{BPFMO}}(t)$ has no dependence on the initial state in the roll.
%It is, however, not easy to relate the time constants ($\tau_{1}$, $\tau_{2}$ and $\tau_{3}$) in Table~\ref{tab:timeconst}, which are usually reported for experimental data analysis~\cite{Martiskainen2009,Martiskainen2012a}, to the time dependent rates in Fig.~\ref{fig:rate}.

\section{Conclusion}
The green sulfur bacteria are thought to be an incredibly efficient light processing machine (\emph{cf.} purple bacterium in Ref.~\cite{Hildner2013}). We studied this system by investigating from all-atoms and a top to bottom approach (\emph{cf.} Ref.~\cite{Linnanto2013}). The excitation energy transfer route was taken from the chlorosome to the reaction center via the baseplate and FMO, under different initiating conditions. Analysis of the atomistic model indicates that resonant energy transfer is maximized given the multichromophoric excitonic coupling which is due to the molecular arrangements of these parts: the green sulfur bacteria are assembled to be most conducive towards efficient excitation energy transfer within the F$\ddot{\mathrm{o}}$rster energy transfer regime. 
It was further shown that whether the initial excitations are important in the energy funneling process. Though, the results differ qualitatively within a short time limit (500\,fs). None of these scenario's, however, adversely affect the efficiency of energy transfer and the results converge within the overall timescale  (10\,ps)~\cite{Martiskainen2009,Martiskainen2012a}. Thus the mechanism is robust to initial conditions, including varying temperatures. This is due mainly to  the fast internal exciton dynamics of the chlorosome, which is also observed in Refs.~\cite{Fujita2012,Fujita2013}.
Furthermore our measure of cooperativity quantifies this and indicates a preference (again regardless of initial conditions) to the polarization in the xy-plane (cross section of the chlorosome), which enhances the excitonic coupling strengths between the exciton states of the chlorosome and the baseplate. We suggest a multichromophoric effect may prevail over the absence of proximity by exploiting the symmetry in parts of the model. This calculation of cooperativity indicates a supertransfer effect inherent in: the green sulfur bacteria, which seems to be especially ``tuned'' towards thriving under low light conditions by making use of molecular aggregates, symmetry and self-assembly to capture light and funnel it to the reaction center aided, not hindered, by a fluctuating environment~\cite{Mohseni2008,Rebentrost2009c}.

Additionally we would like to comment here on the role of the baseplate in the energy
funneling process based on our simulations. In our model study,
the baseplate plays the role of a ``bridge'' allowing the exciton energy to funnel
down to the FMOs from the chlorosome. The presence of the baseplate eases this process; without the baseplate energy transfer would be impeded. Whilst it could be the case that transfer is allowed without the baseplate- under the condition that the FMO's and chlorosomes be positioned close enough for F$\ddot{\mathrm{o}}$rster energy transfer- our results indicates that the baseplate offers a preferred route. The baseplate receives the exciton quickly as shown in Fig.~\ref{fig:popen} and releases the exciton to the FMO complexes steadily. Since the chlorosome has a
relatively large reorganization energy, which implies strong exciton-phonon couplings to the bath, compared to those of the baseplate and FMO complexes, the exciton could be lost to the environment if it is able to stay in the chlorosome for too long a time. Thus, we would like to introduce the idea of the baseplate as a biological  ``exciton capacitor''.  It seems to be suitably designed for this purpose, making sure the route of the exciton is directed, by receiving the exciton from the chlorosome quickly, keeping the exciton from leaking to the surrounding environment, and supplying it to the FMO. It does so by providing appropriate excitonic sites, via chromophoric pigments, held in a unique and protein scaffold made of amphiphillic units that cross two very different dielectric boundaries (the interim gap between dry lipid chlorosomes and the more watery region at the FMOs) in a near perfect 2D lattice form in analogy to an actual capacitor (condenser) but made of soft materials. 

Our model study depends on many undetermined parameters, such as the site energies of the baseplate, distance between the antenna units and the spectral density of the baseplate. 
Also, the structure of the chlorosome is still arguable~\cite{Holzwarth1994,Frese1997,Psencik2004,Linnanto2008,
Ganapathy2009,Ganapathy2012,Tang2013}. 
However, our study shows characteristic time constants that fall within sub-100\,fs-sub-100\,ps and agree with experimental observations~\cite{Martiskainen2009,Martiskainen2012a} (or see Table~\ref{tab:timeconst}). 

%An even more fully comprehensive study using quantum mechanics/molecular mechanics (QM/MM) calculations for the whole system with various other types of chlorosome models~\cite{Holzwarth1994,Frese1997,Psencik2004,Linnanto2008,Ganapathy2012}, including the reaction center, where the electron transfer occurs is in progress in our group.

\begin{acknowledgments}
J. H. thanks Christoph Kreisbeck for the verification of numerical stability in the simulation and discussion about the spectral density.
J. H., S. V., T. F. and A. A.-G. acknowledge support from the Center for Excitonics, an Energy Frontier Research Center funded by the US Department of Energy, Office of Science and Office of Basic Energy Sciences under award DE-SC0001088.
J. C. B. acknowledges support from Welcome Trust UK.
S. K. S. and A. A.-G. also acknowledge Defense Threat Reduction Agency grant HDTRA1-10-1-0046.
Further, A. A.-G. is grateful for the support from Defense Advanced Research Projects Agency grant N66001-10-1-4063, and the Corning Foundation for their generous support.

\end{acknowledgments}

\newpage
\appendix
\renewcommand\thefigure{\thesection.\arabic{figure}}  
\setcounter{figure}{0} 
\section{Supporting Information}
\subsection{Method}
Herein, the method that we use for the exciton propagation and 
computational details is presented.   

Single exciton dynamics is described by Redfield 
quantum master equation (QME), which includes coherent, dephasing and relaxation processes~\cite{Louisellbook,Breuerbook,Adolphs2006,Rebentrost2009,Rebentrost2009b}. The complete system-bath Hamiltonian $\hat{H}$ of the light-harvesting apparatus is composed of three parts, \emph{i.e.}  $\hat{H}=\hat{H}_{\mathrm{S}}+\hat{H}_{\mathrm{SB}}+\hat{H}_{\mathrm{B}}$. 
$\hat{H}_{\mathrm{S}}$ is the system part of Hamiltonian, which describes the local excitations  of $N_{\mathrm{S}}$ bacteriochlorophylls (BChls) and the 
interaction between them. The corresponding tight-binding Hamiltonian is given in a site basis $\vert m \rangle$ as, 
\begin{align}
\hat{H}_{\mathrm{S}}=\sum_{m=1}^{N_{\mathrm{S}}} \epsilon_{m}\vert m \rangle\langle m \vert+\sum_{n<m}^{N_{\mathrm{S}}}V_{mn}(\vert m \rangle\langle n \vert+\vert n \rangle\langle m \vert) 
,~\label{eq:systemhamiltonian}
\end{align}
where $\epsilon_{m}$ is a single excitation energy of the two-level system in site $m$ and $V_{mn}$ is the point dipole interaction energy between the site $m$ and $n$. $\hat{H}_{\mathrm{SB}}$ is the linear coupling term between the system (BChls) and the bath (proteins) coordinate $\hat{q}_{m}$ with a coupling strength $k_{m}$, \emph{i.e.} $\hat{H}_{\mathrm{SB}}=\sum_{m=1}^{N_{\mathrm{S}}} k_{m}\hat{q}_{m}\vert m \rangle\langle m \vert$. $\hat{H}_{\mathrm{B}}$ is the bath Hamiltonian of multidimensional quantum harmonic oscillators. 
Within the secular (numerical degeneracy within 1 cm$^{-1}$) approximation and Markov limit, the Redfield QME of the (reduced) density operator $\hat{\rho}_{\mathrm{S}}$ in the exciton basis is given as follow~\cite{Louisellbook}, for the diagonal and the off-diagonal elements, respectively,  
\begin{align}
&\dot{\rho}_{\mathrm{S},KK}(t)=
\sum_{M}\left(\gamma_{KM}\rho_{\mathrm{S},MM}(t)-\gamma_{MK}\rho_{\mathrm{S},KK}(t)\right)
,\\
&\dot{\rho}_{\mathrm{S},KL}(t)=\nonumber \\
&\left(
-\mathrm{i}\Delta E_{KL}+\gamma_{KL}'
-\tfrac{1}{2}\sum_{M}(\gamma_{MK}+\gamma_{ML})
\right)
\rho_{\mathrm{S},KL}(t),
\label{eq:QME}
\end{align}
where we additionally ignored the Lamb shift.   $\hat{H}_{\mathrm{S}}^{\mathrm{EX}}$ is the exciton Hamiltonian, which is in a diagonal form with the exciton eigenvalues $E_{M}$, 
that is $\hat{H}_{\mathrm{S}}=
\hat{C}\hat{H}_{\mathrm{S}}^{\mathrm{EX}}\hat{C}^{\dagger}$. 
%
%Lindblad operator is given in the exciton basis $\vert M\rangle=\sum_{m}C_{mM}\vert m \rangle$, 
%\begin{align}
%\hat{L}_{\alpha}=\hat{L}_{MN}=\sqrt{\gamma_{MN}}\vert M\rangle\langle N\vert
%,
%\end{align}
$\gamma_{MN}$ is the exciton transition rate between the corresponding 
exciton states $\vert M\rangle$ and $\vert N\rangle$. $\gamma_{MN}$ is calculated with the exciton eigenvectors and spectral density $J_{m}(\Delta E_{MN}/\hbar)$ at the transition energy $\Delta E_{MN}=E_{M}-E_{N}$ and the  reciprocal temperature $\beta$ (see Ref.~\cite{Valleau2012b} for the definition), 
\begin{align}
&\gamma_{MN}(\tfrac{\Delta E_{MN}}{\hbar};\beta)=\nonumber \\ 
&\pi(1+\coth(-\tfrac{\beta\Delta E_{MN}}{2})) \sum_{m}^{N_{\mathrm{S}}}\vert C_{mM}\vert^2\vert C_{mN}\vert^2
J_{m}(-\tfrac{\Delta E_{MN}}{\hbar}), \\
&\gamma_{MN}'(\beta)= 
\frac{2\pi}{\beta} \sum_{m}^{N_{\mathrm{S}}}\vert C_{mM}\vert^2\vert C_{mN}\vert^2
\frac{\mathrm{d}J_{m}(\omega)}{\mathrm{d}\omega}\big\vert_{\omega=0}  
.\label{eq:gammamatrix}
\end{align}
$\gamma_{MN}$ satisfy the detailed balance condition, accordingly, $\gamma_{MN}=\gamma_{NM}\exp(\beta\Delta E_{MN})$.  

\subsection{Spectral density}
We present the spectral densities we used in our simulations. 
\begin{figure}[!htb]
\centerline{\includegraphics[width=1\columnwidth]{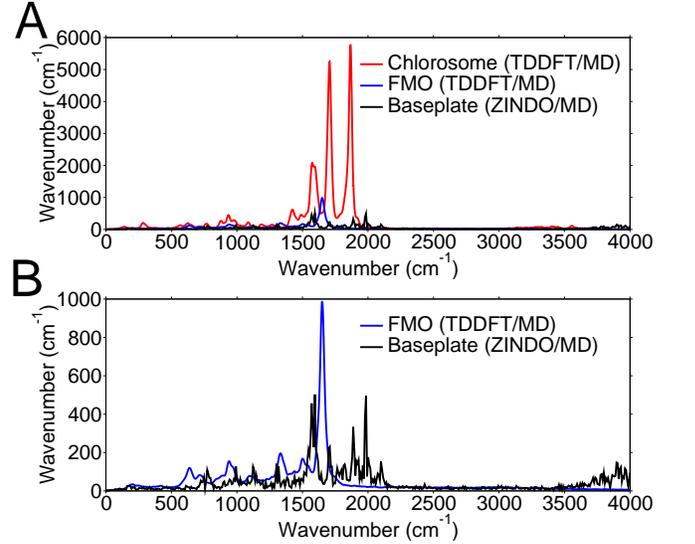}}
\caption{{\bf Spectral density.} 
{\bf A}. Spectral densities were obtained individually for each photosynthetic unit from our previous works. The spectral density of 
the chlorosome (red line) and the FMO complexes (blue line) are collected from Fujita et al.~\cite{Fujita2012} and Shim et al.~\cite{Shim2012}. 
We present here the spectral density of the baseplate obtained from ZINDO/MD calculations only to compare it with the spectral density of the FMO in the low frequency region. 
The ZINDO/MD spectral density of the baseplate is not used in the 
current work. Instead, we used the spectral density of the FMO complex 
as the one of the baseplate because we want to use spectral densities from the same methods. 
We note here again that the spectral densities are taken from the previous works~\cite{Fujita2012,Valleau2012b} not from the quantum mechanics / molecular mechanics calculations of the current model system (chlorosome+baseplate+FMO).    
{\bf B}. The figure is magnified to compare the spectral densities of the baseplate and the FMO complex. We found that the spectral density of the baseplate is not too different from the one of FMO complex in the low frequency domain ($<500$ cm$^{-1}$), which is mainly responsible for the 
exciton transfer between the chlorosome and the baseplate.  
}
\end{figure}

\subsection{Baseplate lattice model}
We present the lattice model of the baseplate used in the current study.  
\begin{figure}[!htb]
\centerline{\includegraphics[width=1\columnwidth]{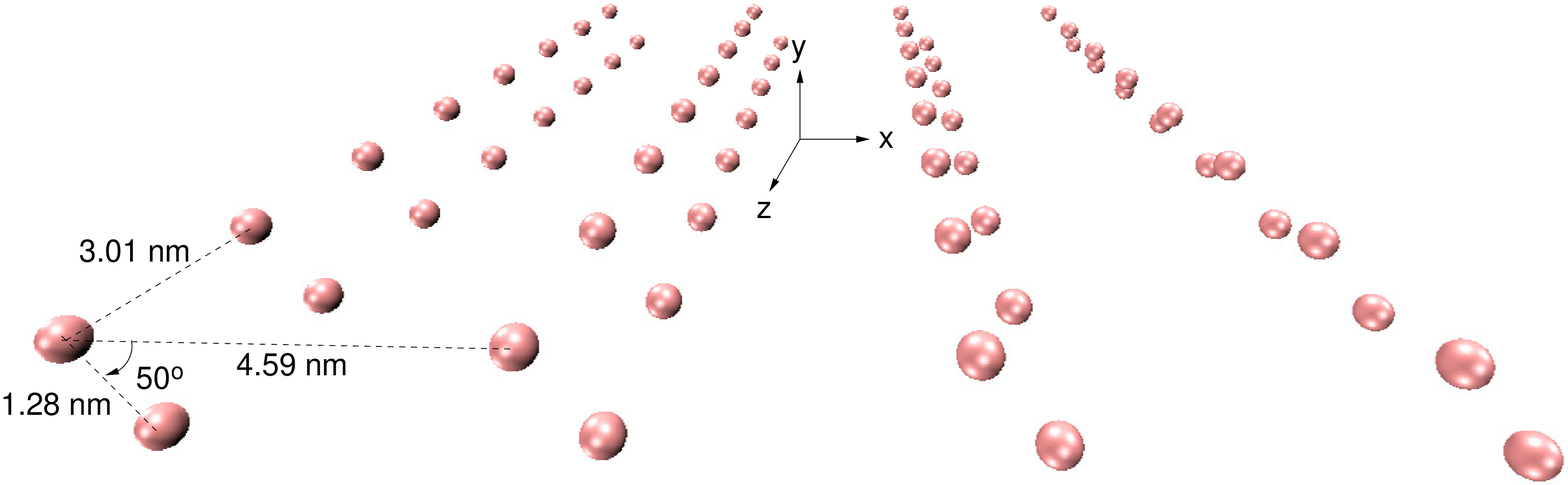}}
\caption{{\bf Baseplate lattice model.} 
The sphere represent the position of the Mg atom of the pigment. 
The lattice model has two layers. Each layer has 32 pigments. 
Two different transition dipole moments are used for the pigments in the different layers. The dipole moment vectors are $\mu_\mathrm{top}=\sqrt{30}(0.2795,0.7484,0.5982)$ and
$\mu_\mathrm{bottom}=\sqrt{30}(0.2533,0.1607,-0.9533)$, respectively for the top and bottom layers. The direction of the dipole moment is approximated by the N1--N3 vector of the BChls. $y$ direction faces to the chlorosome and $-y$ does to the FMO complexes in the model.   
}
\end{figure}

\subsection{Exciton dynamics}
We present here the exciton dynamics, which are not shown in the article.  
\begin{figure}[!htb]
\centerline{\includegraphics[width=1\columnwidth]{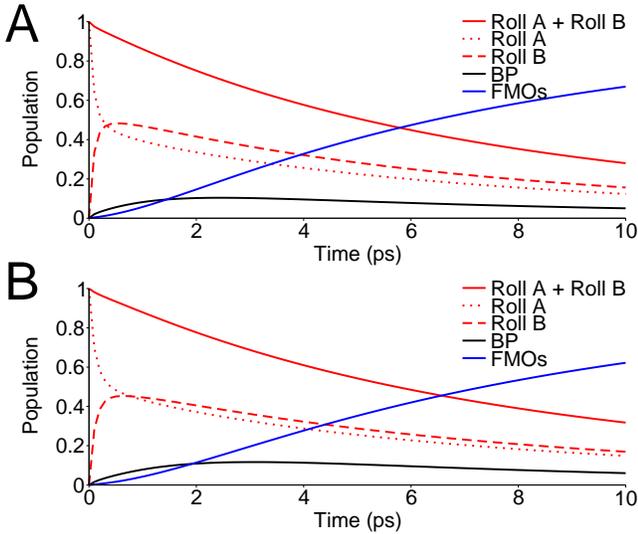}}
\caption{{\bf Exciton dynamics at 150 K and 77 K.} 
{\bf A.} The exciton dynamics at 150 K with the brightest delocalized initial state of the Roll A.   
{\bf B.} The exciton dynamics at 77 K with the brightest delocalized initial state of the Roll A.   
}
\end{figure}

\begin{figure}[!htb]
\centerline{\includegraphics[width=1\columnwidth]{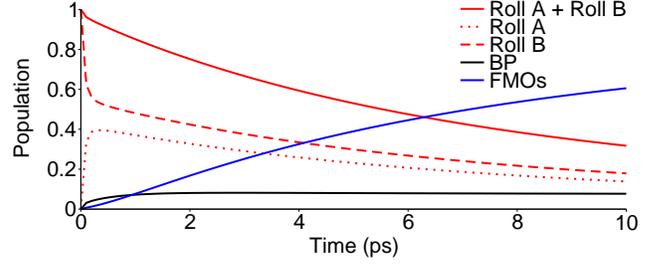}}
\caption{{\bf Exciton dynamics at 300 K with the brightest delocalized initial state of Roll B.} 
The exciton dynamics is obtained with the brightest delocalized initial state of Roll B. 
}
\end{figure}

\begin{figure}[!htb]
\centerline{\includegraphics[width=1\columnwidth]{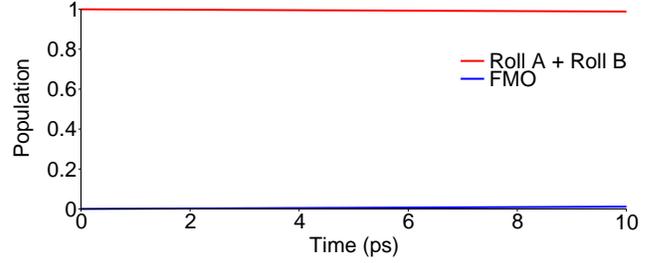}}
\caption{{\bf Exciton dynamics without the baseplate.} 
We test the exciton dynamics without the baseplate. The result is from a single trajectory calculation (the ensemble average is not taken).  
}
\end{figure}

\subsection{Static disorder}
Here, we present the effect of the static disorder. 
\begin{figure}[!htb]
\centerline{\includegraphics[width=1\columnwidth]{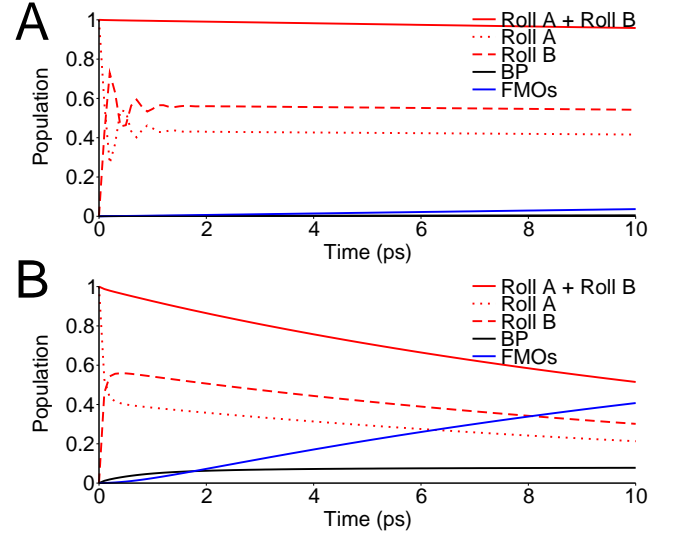}}
\caption{{\bf Exciton dynamics with and without the static disorder.} 
The dynamics were obtained at 300 K with the brightest delocalized initial state of the Roll A.   
{\bf A.} The exciton dynamics without the static disorder. 
{\bf B.} The exciton dynamics with the static disorder. The results are from single trajectory calculations (the ensemble average is not taken). 
\label{fig:singleed}}
\end{figure}

\begin{figure}[!htb]
\centerline{\includegraphics[width=1\columnwidth]{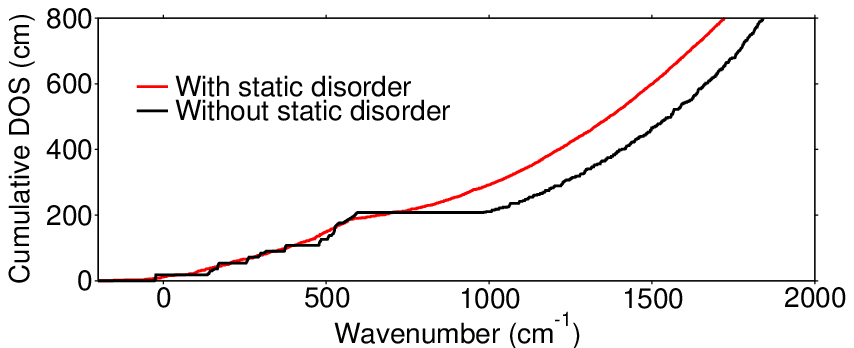}}
\caption{{\bf Cumulative density of states (DOS) with and without the static disorder.}\\ 
Cumulative-DOS($\omega$)=$\int_{0}^{\omega}\mathrm{d}\omega$DOS($\omega$). 
\label{fig:singleds}}
\end{figure}
There is almost no exciton transfer when the static disorder is not considered (Fig.~\ref{fig:singleed}{\bf A}) because the spectral broadening causes the spectral overlap between the antenna units. The single trajectory 
with the static disorder is already similar to the 1000 ensemble averaged result. This is because there is no exciton states in 500--1000 cm$^{-1}$ 
when the static disorder (Fig.~\ref{fig:singleds}) is not considered. This energy domain is important for the exciton transfer between the chlorosome and the baseplate.    

\newpage

%\bibliographystyle{pnas2009}
%\bibliographystyle{apsrev4-1}
%\bibliographystyle{pnas}
%\bibliography{pnas_huh2013}
%\begin{thebibliography}{94}
%\end{thebibliography}

\end{document}